\documentclass[showpacs,preprintnumbers,amsmath,amssymb]{revtex4}

\usepackage{epsfig}
\usepackage{graphicx}% Include figure files
\usepackage{dcolumn}% Align table columns on decimal point
\usepackage{bm}% bold math
\usepackage[active]{srcltx}
\newcommand{\be}{\begin{equation}}

\newcommand{\ee}{\end{equation}}
\newcommand{\bea}{\begin{eqnarray}}
\newcommand{\eea}{\end{eqnarray}}

\begin{document}

\preprint{}

\title{Granger causality and the inverse Ising problem}
\date{\today}
%\author{D.~Marinazzo}\email{daniele.marinazzo@ba.infn.it}
\author{Mario~Pellicoro}\email{mario.pellicoro@ba.infn.it}
\author{Sebastiano~Stramaglia}\email{sebastiano.stramaglia@ba.infn.it}
\affiliation{Dipartimento  Interateneo di Fisica, Universit\`a di
Bari, I-70126 Bari, Italy,} \affiliation{TIRES-Center of Innovative
Technologies for Signal Detection and Processing, Universit\`a di
Bari,  Italy,} \affiliation{I.N.F.N., Sezione di Bari, I-70126 Bari,
Italy}

\date{\today}% It is always \today, today,

\begin{abstract}
The inference of the couplings of an Ising model with given means
and correlations is called {\it inverse Ising problem}. This
approach has received a lot of attention as a tool to analyze neural
data. We show that autoregressive methods may be used to learn the
couplings of an Ising model, also in the case of asymmetric
connections and for multi-spin interactions. We find that, for each
link, the linear Granger causality is two times the corresponding
transfer entropy (i.e. the information flow on that link) in the
weak coupling limit. For sparse connections and a low number of
samples, the $\ell_1$ regularized least squares method is used to
detect the interacting pairs of spins. Nonlinear Granger causality
is related to multispin interactions.
\end{abstract}

\maketitle

\section{Introduction}
A large of neurons is involved in any computation, and the presence
of non-trivial correlations makes understanding  the mechanisms of
computation in brain a difficult challenge \cite{rieke}. The
simplest model for describing multi-neuron spike statistics is the
pairwise Ising model \cite{sch,shlens}. The inference of the
couplings of an Ising model from data, the {\it inverse Ising
problem}, has recently attracted attention, see \cite{hertz} where
data from a simulated cortical network were considered; the general
idea is to find an Ising model with the same means and pairwise
correlations as the data. Several approximate methods can be used,
like that by Sessak and Monasson \cite{sm} or the inversion of TAP
equations \cite{tap}: equal-time correlations from data are used in
those methods \cite{roudi}. The flow of information between spins is
related, instead, to correlations at different times between
variables: it can be expected that measuring the information flow
between spins one may improve the estimate of couplings from data.

Two major approaches are commonly used to estimate the information
flow between variables, transfer entropy \cite{schreiber} and
Granger causality \cite{granger}. Recently it has been shown that
for Gaussian variables Granger causality and transfer entropy are
entirely equivalent as the following relation holds: {\it Granger
causality  = 2 Transfer Entropy}. This result provides a bridge
between  autoregressive and information-theoretic approaches to
causal inference from data \cite{seth}.

The purpose of this work is to explore the use of Granger causality
to learn Ising models from data. The {\it inverse Ising problem} is
here seen as belonging to the more general frame of the inference of
dynamical networks from data, a topic which has been studied in
recent papers \cite{yu1,sauer,noipre,materassi}: its relevance is
due to the fact that dynamical networks \cite{barabasi} model
physical and biological behavior in many applications \cite{bocca}.
We show that for weak couplings, the linear Granger causality of
each link is two times the corresponding transfer entropy, also for
Ising models: this occurrence justifies the use of autoregressive
approaches to the inverse Ising problem. In the same limit, for each
link, the following relation exists between the coupling ($J$) and
the causality ($\delta$): $\delta=J^2$. In the case of limited
samples, Granger causality gives poor results:  almost all the
connections are not assessed as significative for low number of
samples. In these cases, we propose the use of $\ell_1$ least
squares method \cite{tib}, a penalized autoregressive approach
tailored to embody the sparsity assumption, to recover the non
-vanishing connections of a sparse Ising model; as expected the
$\ell_1$ approach outperforms Granger causality in this case.
Finally we show that nonlinear Granger causality is related to
multi-spin interactions.

The paper is organized as follows. In the next section we briefly
recall the notions of Granger causality and transfer entropy, and we
also describe the Ising models that we use for simulations. In
section III we describe our results on fully connected models,
sparse Ising models and models with higher order spin interactions.
Some conclusions are drawn in section IV.

\section{Granger causality and Transfer entropy}
In this section we review the notions of  Granger causality analysis
and transfer entropy. We also discuss the application of these
methods to binary time series arising from Ising models.
\subsection{Granger causality}
Granger causality has become the method of choice to determine
whether and how two time series exert causal influences on each
other \cite{hla}. It is based on prediction: if the prediction error
of the first time series is reduced by including measurements from
the second one in the linear regression model, then the second time
series is said to have a causal influence on the first one
\cite{dingpla}. The estimation of linear Granger causality from
Fourier and wavelet transforms of time series data has been recently
addressed \cite{ding-prl}. The nonlinear generalization of Granger
causality has been developed by a kernel algorithm which embeds data
into a Hilbert space, and searches for linear Granger causality in
that space \cite{noiprl}; the embedding is performed implicitly, by
specifying the inner product between pairs of points \cite{shawe},
and a statistical procedure is used to avoid over-fitting.

Quantitatively, let us consider $n$ time series $\{x_\alpha
(t)\}_{\alpha =1,\ldots,n}$ \cite{nota}; the lagged state vectors
are denoted
$$X_\alpha (t)= \left(x_\alpha (t-m),\ldots,x_\alpha (t-1)\right),$$
$m$ being the window length.  Let $\epsilon \left(x_\alpha |{\bf
X}\right)$ be the mean squared error prediction of $\bf{x}_\alpha$
on the basis of all the vectors ${\bf X}$ (corresponding to the
kernel approach described in \cite{noiprl}): $\epsilon
\left(x_\alpha |{\bf X}\right)$ is equal to
$1-\tilde{\bf{x}}_\alpha^\top\tilde{\bf{x}}_\alpha$, where
$\tilde{\bf{x}}_\alpha$, the predicted values of $\bf{x}_\alpha$
using ${\bf X}$, is the projection of $\bf{x}_\alpha$ on a suitable
space $H$. The prediction of $\bf{x}_\alpha$ on the basis of all the
variables but $X_\beta$, $\epsilon \left(x_\alpha |{\bf X}\setminus
X_\beta\right)$, corresponds instead to the projection on a space
$H'$ with $H=H'\oplus H^\perp$. $H^\perp$ represents the information
that one gains from the knowledge of $X_\beta$. The multivariate
Granger causality index $\delta (\beta \to \alpha )$ is defined as
the (normalized) variation of the error in the two conditions, i.e.
\begin{equation}\label{delta}
\delta (\beta \to \alpha )={\epsilon \left(x_\alpha |{\bf
X}\setminus X_\beta\right)-\epsilon \left(x_\alpha |{\bf
X}\right)\over \epsilon \left(x_\alpha |{\bf X}\setminus
X_\beta\right)}.
\end{equation}
Note that the numerator, in the equation above, coincides with the
projection of  $\bf{x}_\alpha$ on $H^\perp$: as described in
\cite{noipre}, one may write \begin{equation}\label{somma}\delta
(\beta \to \alpha )=\sum_{i=1}^m r_i^2,\end{equation} where $r_i$
are suitable Pearson's correlations. By summing, in equation
(\ref{somma}), only over significative correlations,  a {\it
filtered} linear Granger causality index is obtained which measures
the causality without further statistical test.

In \cite{ancona} it has been shown that not all the kernels are
suitable to estimate causality. Two important classes of kernels
which can be used to construct nonlinear causality measures are the
{\it inhomogeneous polynomial kernel} (whose features are all the
monomials, in the input variables, up to the $p-th$ degree; $p=1$
corresponds to linear Granger causality) and the {\it Gaussian
kernel}. Note that in \cite{seth} a different index of causality is
adopted:
\begin{equation}\label{delta1}
\Delta (\beta \to \alpha )= \log{\epsilon \left(x_\alpha |{\bf
X}\setminus X_\beta\right)\over \epsilon \left(x_\alpha |{\bf
X}\right)}=-\log{\left[1-\delta (\beta \to \alpha )\right]};
\end{equation}
$\Delta$ and $\delta$ coincide at small $\delta$. The choice of the
window length $m$ is usually done using the standard
cross-validation scheme \cite{10fold}; as in this work we know how
data are generated, here we use $m=1$.

The formalism of Granger causality is constructed under the
hypothesis that time series assume continuous values $x_\alpha$. In
recent papers the application of Granger causality to data in the
form of phases has been considered \cite{noipla,smirnov}. Even
though there is not theoretical justification in the case of binary
variables, here we apply the formalism of Granger casuality to $n$
binary time series $\{\sigma_\alpha (t)=\pm 1\}_{\alpha
=1,\ldots,n}$, by substituting
$$x_\alpha (t) \to \sigma_\alpha (t)$$ and $$X_\alpha (t) \to  \sigma_\alpha (t-1)=\Sigma_\alpha (t);$$
in this work we  justify the application of Granger causality to
binary time series in terms of its relation with transfer entropy.

\subsection{Transfer entropy}
Using the same notation as in the previous subsection, the transfer
entropy index $T_E (\beta \to \alpha )$ is given by \cite{schreiber}
\begin{equation}\label{tau}
T_E (\beta \to \alpha )=\int dx_\alpha \int d{\bf X} \; p\left(
x_\alpha,{\bf X}\right) \log{p\left(x_\alpha |{\bf X}\setminus
X_\beta\right)\over p\left(x_\alpha |{\bf X}\right)},
\end{equation}
and measures the flow of information from $\beta$ to $\alpha$. For
Gaussian variables it has been shown \cite{seth} that causality is
determined by the transfer entropy and $\Delta= 2 T_E$; hence
$\delta = 1-e^{-2T_E}$ for Gaussian variables. The probabilities
$p$'s, in (\ref{tau}), must be estimated from data using techniques
of nonlinear time series analysis \cite{kantz}. In the case of
binary variables $\{\sigma_\alpha =\pm 1\}$, the number of
configurations is finite and the integrals in (\ref{tau}) become
sums over configurations; the probabilities can be estimated as
frequencies in the data-set at hand.  Therefore
\begin{equation}\label{te}
T_E (\beta \to \alpha )=\sum_{\sigma_\alpha =\pm 1} \sum_{\Sigma_1
=\pm 1}\cdots \sum_{\Sigma_n =\pm 1}  p\left( \sigma_\alpha,{\bf
\Sigma}\right) \log{p\left(\sigma_\alpha ,{\bf \Sigma}\setminus
\Sigma_\beta\right)\; p\left({\bf \Sigma}\right) \over
p\left(\sigma_\alpha ,{\bf \Sigma}\right)\;p\left({\bf
\Sigma}\setminus \Sigma_\beta\right)},
\end{equation}
where $p\left({\bf \Sigma}\right)$ is the fraction of times that the
configuration ${\bf \Sigma}$ is observed in the data set, and
similar definitions hold for the other probabilities. We remark that
the number of configurations increases exponentially as the number
of spins grows, hence the direct evaluation of (\ref{te}) is
feasible only for systems of small size.
\subsection{Ising models}
The binary time series analyzed in this work are generated by
parallel updating of Ising variables $\{\sigma_\alpha \}_{\alpha
=1,\ldots,n}$:
\begin{equation}\label{update}
p\left(\sigma_\alpha(t)=+1 |{\bf \Sigma}(t)\right)={1\over
1+e^{-2h_\alpha (t)}},
\end{equation}
where  the local fields are given by
\begin{equation}\label{delta}
h_\alpha (t)=\sum_{\beta=1}^n J_{\alpha \beta} \sigma_\beta (t-1)
\end{equation}
with couplings $J_{\alpha \beta}$. Starting from a random initial
configuration of the spin, equations (\ref{update}) are iterated
and, after discarding the initial transient regime, $N$ consecutive
samples of the system are stored for further analysis.

\section{Analysis of Ising models}
\subsection{Fully connected  models}

\begin{figure}[ht]
\includegraphics[width=10cm]{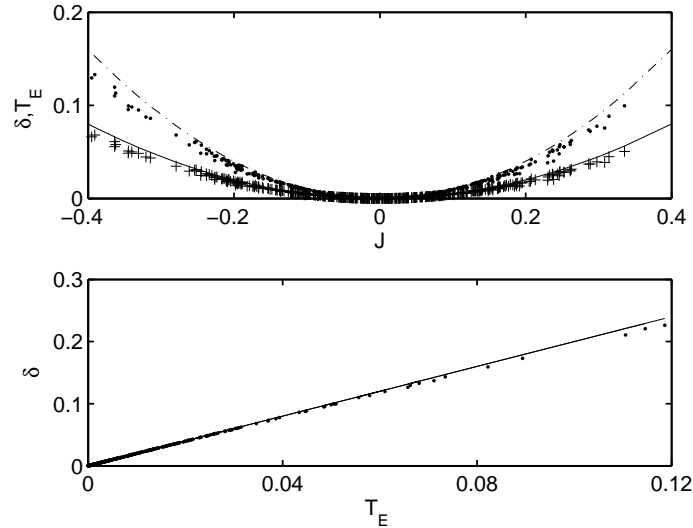}
\caption{{\rm (Top) The estimates of linear Granger causality and
transfer entropy, for each link,  are plotted versus the coupling
$J$. The points, corresponding to 15 realizations of the couplings
of six-spins Ising systems with $J_0$ ranging in $[0.1,0.2]$, are
displayed (the two quantities are estimated over samples of $N=10^6$
length). The curves are the quadratic expansions at weak coupling:
$\delta= J^2$ (dashed-dotted line) and $T_E =J^2/ 2$ (continuous
line). (Bottom) The same points are displayed in the $\delta-T_E$
plane, showing that $\delta = 2 T_E$ at weak coupling.
\label{f1}}}\end{figure}

In order to generalize Granger causality to discrete variables, we
consider the regression function  of (\ref{update}). For weak
couplings the conditional expectation (which coincides with the
regression function \cite{papoulis}) can be written
\begin{equation}\label{condexp}
\langle \sigma_\alpha \rangle_{|\Sigma}=\sum_{s=\pm 1} s
\;p\left(\sigma_\alpha =s | {\bf \Sigma}\right) =\mbox{tanh}
(h_\alpha)\sim \sum_{\beta=1}^n J_{\alpha \beta} \Sigma_\beta,
\end{equation}
and is a linear function of the couplings. Therefore the linear
causality is $\delta (\beta \to \alpha ) \sim J_{\alpha \beta}^2$ at
the lowest order in $J$'s. Analogously, expanding eq.(\ref{te}) at
the lowest order, the transfer entropy reads $T_E (\beta \to \alpha
) \sim J_{\alpha \beta}^2/ 2$. This means that, for low couplings,
the value of $J$, for any given link, determines both the transfer
entropy and the linear Granger causality for that link, and the two
quantities differ only by the factor $2$. The same relation, proved
in the Gaussian case, holds also for Ising models at weak couplings.
Being related to the transfer entropy, Granger causality thus
measures the information flow for these systems, and this justifies
the use of Granger causality methods for Ising systems. Synaptic
couplings are directed, so $J_{\alpha \beta}$ is not in general
equal to $J_{\beta \alpha }$ (the equilibrium Ising model requires
symmetric couplings \cite{kadanoff}). Therefore we consider an
asymmetric system of spins with couplings $J_{\alpha \beta}$ chosen
at random from a normal distribution with zero mean and standard
deviation $J_0$; no self interactions are assumed ($J_{\alpha\alpha
} =0$). In figure (\ref{f1}) we report the plot of the numerical
estimates of linear causality and transfer entropy, as a function of
$J$, for several realizations of the couplings and for some values
of $J_0$; the simulations confirm that for low couplings a one-one
correspondence exists between causality and transfer entropy.

\begin{figure}[ht]
\includegraphics[width=10cm]{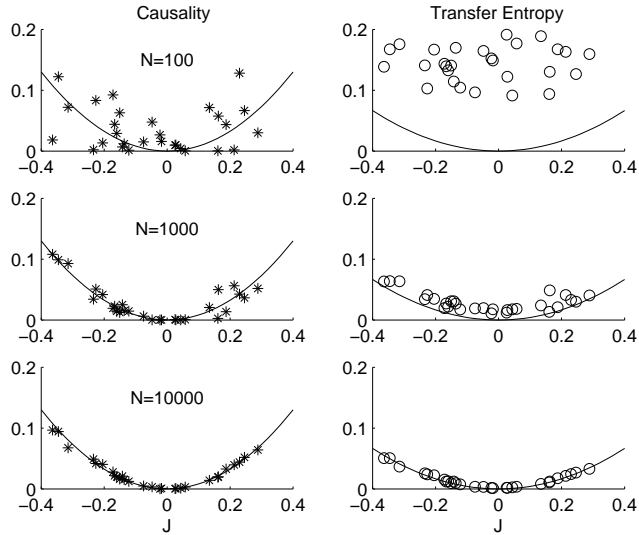}
\caption{{\rm For a typical asymmetric Ising model of six spins and
$J_0 = 0.2$, the linear  Granger  causality  (right) and the
transfer entropy (left) are depicted for each link, as a function of
its coupling $J$,  for $N=100$ (top), $1000$ (middle) and $10000$
(bottom) samples. The continuous curves represent the {\it true}
values (obtained by fitting the points in fig.
1).\label{f2}}}\end{figure}

\begin{figure}[ht]
\includegraphics[width=10cm]{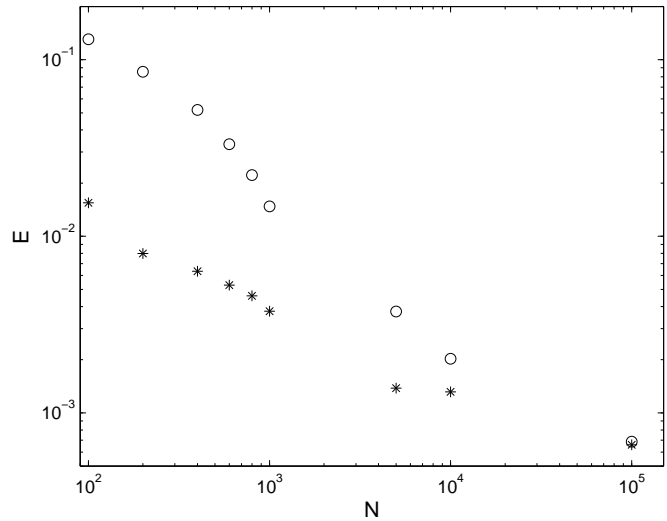} \caption{{\rm
The asymmetric Ising model described in figure 2 is here considered.
Calling $t_a$ the true value of the transfer entropy and $t_b$ its
estimate based on $N$ samples, averaged over 1000 runs of the Ising
system, we define $E=|t_b -t_a|/t_a$. The quantity E, thus obtained,
is here plotted versus N (empty circles). A similar quantity E,
concerning Granger causality, is also plotted
(stars).\label{f3}}}\end{figure}

In figure (\ref{f2}) we depict, as a function of the coupling $J$,
both the linear causality and the transfer entropy in a typical
asymmetric model of six spins with $N=100, 1000$ and $10000$
samples, and $J_0 =0.2$. In figure (\ref{f3}) we depict, as a
function of the number of samples $N$, the difference between the
values of transfer entropy and causality (as estimated on $10^6$
samples) and their estimates based on $N$ samples (the difference is
averaged over 1000 runs of the Ising system): at low $N$ the
estimates of causalities are more reliable than those of transfer
entropy. Moreover, the computational complexity of the estimation of
transfer entropy is much higher than those corresponding to the
evaluation of Granger causality.

\begin{figure}[ht]
\includegraphics[width=10cm]{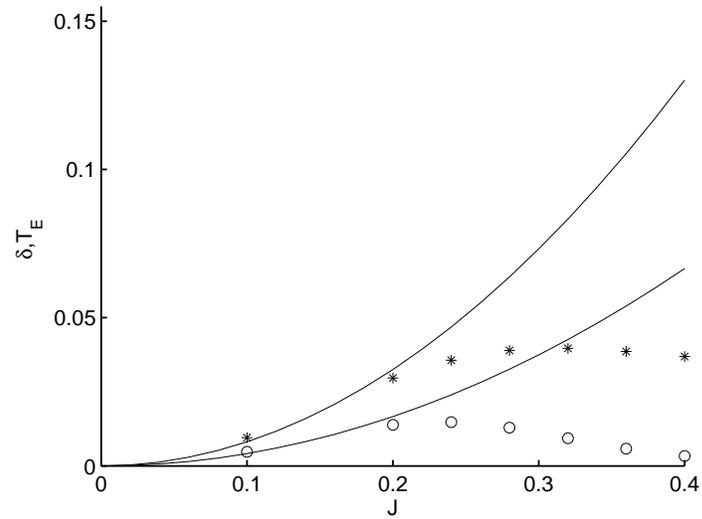}
\caption{{\rm The homogeneous Ising model, with $n=6$ and uniform
couplings $J > 0$, is considered. As a function of $J$, the linear
Granger causality (stars) and the transfer entropy (empty circles)
are depicted versus $J$. Both quantities are the same for all links,
due to symmetry. The two curves are the relations between the
coupling and transfer entropy (and between coupling and causality)
which hold for the asymmetric Ising model (obtained by fitting the
points of fig. 1). \label{f4}}}\end{figure}

\begin{figure}[ht]
\includegraphics[width=10cm]{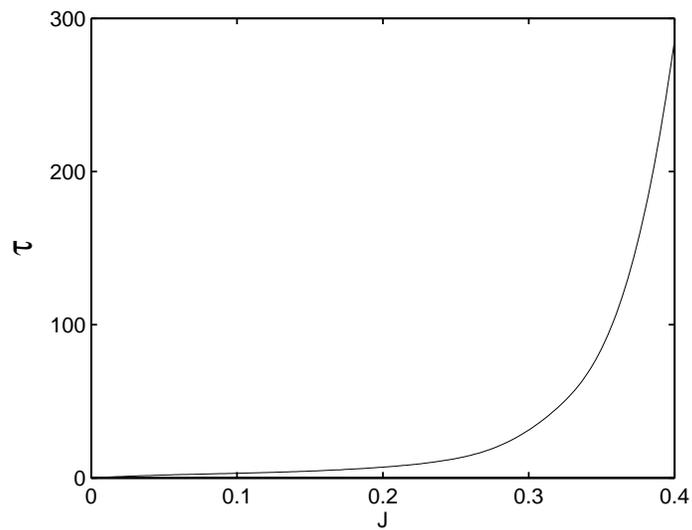}
\caption{{\rm The auto-correlation time of the magnetization for the
homogeneous Ising model analyzed in figure 4.
\label{f5}}}\end{figure}

Another interesting situation is all the couplings being equal to a
positive quantity $J$ (still, without self-interactions). In figure
(\ref{f4}) we depict the Granger causality and the transfer entropy
as a function of $J$ (these quantities are the same for all the
links, due to symmetry). For small values of $J$ both quantities
coincide with the values that correspond to $J$ in the asymmetric
model, and the relation $\delta /T_E =2$ holds. On the other hand,
as $J$ increases, this relation is more and more violated. The
departure  of the ratio $\delta /T_E$ from 2 is connected to the
emergence of feedback effects in the system, see e.g. the
dependency, on $J$, of the auto-correlation time of the
magnetization $n^{-1}\sum_{\alpha =1}^n \sigma_\alpha$ in
fig.(\ref{f5}).
\subsection{Sparse models}
In many applications one may hypothesize that the connections among
variables are sparse. The main goal, in those cases, is to infer the
couplings which are not vanishing, independently of their strengths,
in particular when the number of samples is low. Moreover, in the
case of limited data, Granger causality gives poor results; indeed
almost all connections would not be assessed as significative (for a
given amount a data, only couplings stronger than a critical value
can be recognized by Granger causality \cite{noiprl}).

A major approach to sparse signal reconstruction is the $\ell_1$
regularized least squares method \cite{tib}. Although it has been
developed to handle continuous variables, we will apply this method
to the configurations of Ising models. For each target spin
$\sigma_\alpha$, the vector of couplings $A_{\alpha \beta}$, with
$\beta =1,\ldots,n$, is sought for as the minimizer of
\begin{equation}\label{elleuno}
\sum_{t=1}^N \left(\sigma_\alpha (t) - \sum_\beta A_{\alpha \beta}
\Sigma_\beta (t) \right)^2 + \lambda ||A_{\alpha \beta}||_1,
\end{equation}
where $\lambda > 0$ is a regularization parameter and $||A_{\alpha
\beta}||_1 =\sum_{\beta=1}^n |A_{\alpha \beta}|$ is the $\ell_1$
norm of the vector of couplings. As $\lambda$ is increased, the
number of vanishing couplings in the minimizers increases: $\lambda$
controls the sparsity of the solution \cite{boyd}. The strategy to
fix the value of $\lambda$ we use here is 10-fold cross-validation
\cite{10fold}: the original sample is randomly partitioned into 10
subsamples and, out of the 10 subsamples, a single subsample is
retained as the validation data. The remaining 9 subsamples are used
in (\ref{elleuno}) to determine the couplings $A$; the quality of
this solution is evaluated as the average number of errors on the
validation data.
%$$e={1\over 2n}\sum_{s=1}^{n\over 10}\left(1-\sigma_\alpha
%(s)\;sign\left[\sum_\beta A_{\alpha \beta} \Sigma_\beta
%(s)\right]\right).$$
The cross-validation process is then repeated 10 times (the folds),
with each of the 10 subsamples used exactly once as the validation
data, and the error on the validation data is averaged over the 10
folds. The whole procedure is then repeated as $\lambda$ is varied.
The optimal value of $\lambda$ is chosen as the one leading to the
smallest average error on the validation data.

\begin{figure}[ht]
\includegraphics[width=10cm]{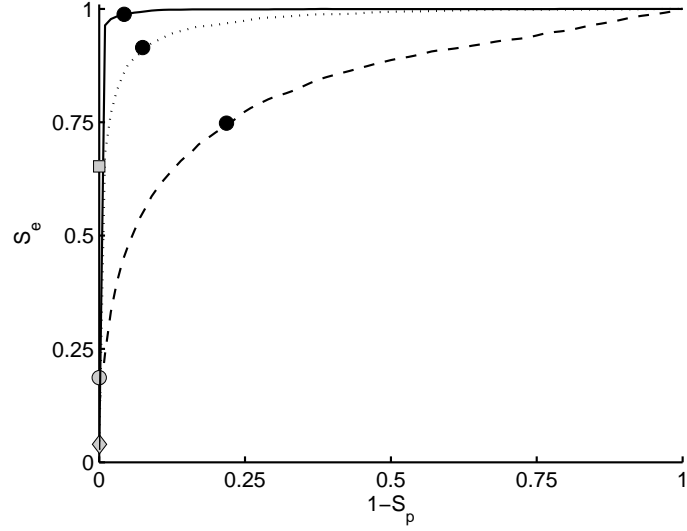}
\caption{{\rm The ROC curves (sensitivity vs 1-specificity) for the
detection of non-vanishing couplings in the sparse system of 30
spins described in the text; the curves correspond  to three values
of the number of samples $N$, 100 (dashed line), 250 (dotted line)
and 500 (continuous line). The stars on these curves represent the
points found by ten-fold cross validation. The other three symbols
are the performances by Granger causality on $N=100$ (empty
diamond), $N=250$ (empty circle), $N=500$ (empty square).
\label{f6}}}\end{figure}

As an example, we simulate a system made of 30 spins constituted by
ten modules of three spins each. The non-vanishing couplings of the
Ising model are given by:

\begin{eqnarray}
\begin{array}{lll}
J(3i-2,3i-1)&=&0.2,\\
J(3i-1,3i)&=&0.2,\\
J(3i,3i-2)&=-&0.2,\\
J(3i,3i-1)&=&0.2,\\
\end{array}
\label{sparse}
\end{eqnarray}
for $i=1,2,\ldots,10$. After evaluating the couplings $A_{\alpha
\beta}$, using the algorithm described in \cite{boyd}, we calculate
the sensitivity (fraction of non-vanishing connections $J$ leading
to non-vanishing couplings $A$) and the specificity (fraction of
vanishing connections $J$ leading to vanishing couplings $A$) as a
function of $\lambda$. The ROC curves (as $\lambda$ is varied, the
ROC curve is sensitivity as a function of 1-specificity \cite{roc})
we obtain, in correspondence to three values of the number of
samples N (100, 250 and 500), are depicted in fig. (\ref{f6}). The
stars on the curves represent the points corresponding to the value
of $\lambda$ found by ten-fold cross validation; these points
correspond to a good compromise between specificity and sensitivity.
The empty symbols, instead, represent the values of sensitivity and
specificity obtained using Granger causality in the three cases; the
specificity by Granger causality is nearly one in all cases, while
the sensitivity is strongly dependent on the number of samples and
goes to zero as $N$ decreases. To conclude this subsection, we have
shown that in the case of low number of samples and sparse
connections  the $\ell_1$ regularized least squares method can be
used to infer the connections in Ising models and outperforms
Granger causality in these situations. We remark that direct
evaluation of the transfer entropy in these cases is unfeasible.

\subsection{Higher order spin interactions}
\begin{figure}[ht]
\includegraphics[width=10cm]{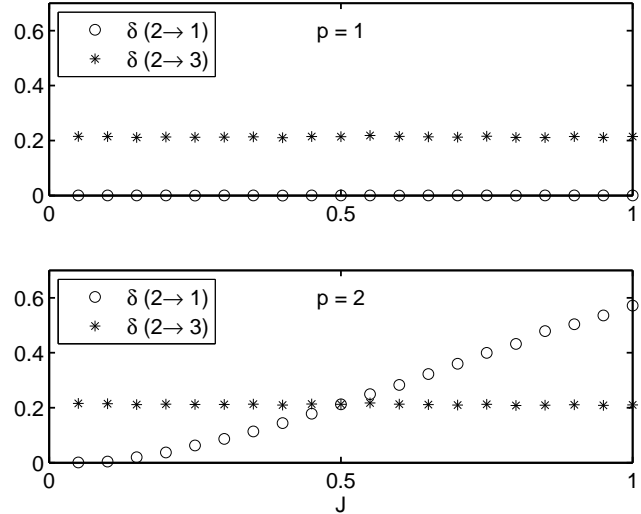}
\caption{{\rm The causalities $\delta (2 \to 1)$ and $\delta (2 \to
3)$  are depicted as a function of $J$ for the three spins system
described in the text. Causalities are estimated using the linear
kernel (top) and the $p=2$ polynomial kernel (bottom).
\label{f7}}}\end{figure}

The case of higher order spin interactions requires use of nonlinear
Granger causality: in the presence of $p-$spins interactions, the
kernel approach \cite{noiprl} with the polynomial kernel of at least
$p-1$ degree is needed. As an example, we consider a system of three
spins with local fields given by:
\begin{eqnarray}
\begin{array}{llc}
h_1 (t)&=&J \sigma_2 (t-1)\sigma_3 (t-1),\\
h_2 (t)&=&0.5\;\sigma_3 (t-1),\\
h_3 (t)&=&0.5\;\sigma_2 (t-1).
\end{array}
\label{sparse}
\end{eqnarray}
In figure (\ref{f7}) we depict the causalities $\delta (2 \to 1)$
and $\delta (2 \to 3)$, as a function of $J$, using the linear
kernel and for the approach with the $p=2$ polynomial kernel. Note
that, due to symmetry, $\delta (3 \to 1)=\delta (2 \to 1)$ and
$\delta (2 \to 3)= \delta (3 \to 2)$; all the other causalities are
vanishing. The linear approach is not able to detect the three spins
interaction, while using the nonlinear approach the interaction is
correctly inferred. We stress that the presence of multispin
interactions is connected to the presence of synergetic variables,
see \cite{redundant} for a discussion about the notions of
redundancy and synergy in the frame of causality.

\begin{figure}[ht]
\includegraphics[width=10cm]{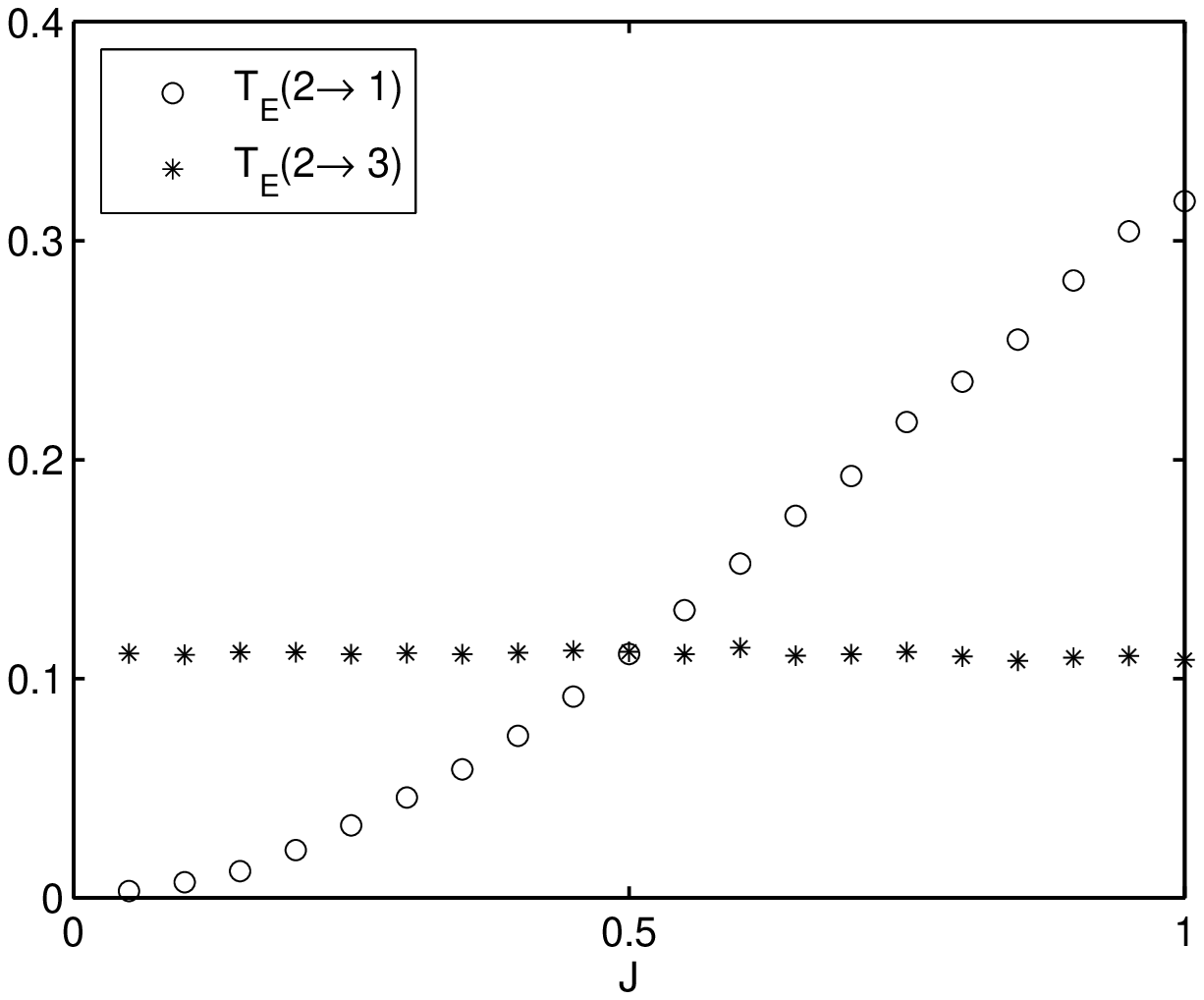}
\caption{{\rm The transfer entropies $T_E (2 \to 1)$ and $T_E (2 \to
3)$ are depicted as a function of $J$ for the three spins system
described in the text. \label{f8}}}\end{figure}

It is interesting to show the performance by transfer entropy on the
same problem, see fig.(\ref{f8}): it correctly detects all the
interactions, and the value of the transfer entropy is again very
close to be half of those from nonlinear Granger causality. We
stress that transfer entropy can be applied without prior
assumptions about the order of the spins interactions. A major
problem in the inference of dynamical networks is the selection of
an appropriate model; in the case of transfer entropy this issue
does not arise, although this advantage may be offset by problems
associated with reliable estimation of entropies in sample.
\section{Conclusions}
We have proposed the use of autoregressive methods to learn  Ising
models from data. Commonly, the formulation of the inverse Ising
problem assumes symmetric interactions and is solved by exploiting
the relations that exist, at equilibrium, between the pairwise
correlations (at equal times) and the matrix of couplings. In the
general case of asymmetric couplings, no equilibrium is reached and
also time delayed correlations among spins should be used to infer
the connections. We have shown that autoregressive approaches can
solve the inverse Ising problem for weak couplings: for each link
$|J_{\alpha \beta}|=\sqrt{\delta (\beta \to \alpha)}$, whilst the
sign of $J$ coincides with the sign of the linear correlation
between $\sigma_\alpha$ and $\Sigma_\beta$. For weak couplings,
Granger causality is proportional to the transfer entropy and
requires less samples, than transfer entropy, to provide a reliable
estimate of the information flow. For sparse connections and low
number of samples, the $\ell_1$ regularized least squares method is
preferable to Granger causality; nonlinear Granger causality is
related to multispin interactions.

The authors thank Amos Maritan and Marco Zamparo (University of
Padova) for valuable discussions.

\begin{thebibliography}{99}
\bibitem{rieke} F. Rieke, D. Warland, R.R. de Ruyter van Steveninck,
and W. Bialek, {\it Spikes: exploring the neural code} (MIT press,
Cambrige, MA, 1997).
\bibitem{sch}E. Schneidman, M.J. Berry, R. Segev, W. Bialek,
Nature {\bf 440}, 1007 (2006).
\bibitem{shlens}J. Shlens, G.D.
Field, J.L. Gauthier, M.I. Grivich, D. Petrusca, A. Sher, A.M.
Litke, E.J. Chichilnisky, J. Neurosci. {\bf 28} 505 (2008).
\bibitem{hertz} Y. Roudi, J. Tyrcha,
J. Hertz, Physical Review E {\bf 79}, 051915 (2009).
\bibitem{sm} V. Sessak and R. Monasson, J. Phys. A {\bf 42}, 055001
(2009).
\bibitem{tap} H.J. Kappen and F.B. Rodriguez, Neural Computation
{\bf 10}, 1137 (1998).
\bibitem{roudi}Y. Roudi, E. Aurell and J.A.
Hertz, Front. Comput. Neurosci. {\bf 3}, 22 (2009).
\bibitem{schreiber} T. Schreiber, Phys. Rev. Lett.
{\bf 85}, 461 (2000)
\bibitem{granger} C.W.J. Granger, Econometrica {\bf 37}, 424 (1969).
\bibitem{seth} L. Barnett, A.B. Barrett, and A.K. Seth, Phys.
Rev. Lett. {\bf 103}, 238701 (2009).
\bibitem{yu1}D. Yu, M. Righero, L. Kocarev, Phys.Rev.Lett. {\bf 97}, 188701 (2006).
\bibitem{sauer}  D. Napoletani, T. Sauer, Phys. Rev. {\bf E 77},
26103 (2008).
\bibitem{noipre}D. Marinazzo, M. Pellicoro and S. Stramaglia, Phys. Rev. E {\bf 77}, 056215 (2008).
\bibitem{materassi} D. Materassi, G. Innocenti, Physica A {\bf 388}, 3866 (2009).
\bibitem{barabasi}A.L. Barabasi, {\em Linked: the new science of networks}. (Perseus Publishing, Cambridge Mass.,
2002). \bibitem{bocca} S. Boccaletti, V.  Latora, Y.  Moreno, M.
Chavez and D.-U. Hwang, Phys. Rep. {\bf 424}, 175 (2006).
\bibitem{tib} R. Tibshirani, J. Roy. Stat. Soc. {\bf B 58}, 267
(1996).
\bibitem{hla} K. Hlavackova-Schindler, M. Palus, M. Vejmelka, J.
Bhattacharya, Physics Reports {\bf 441}, 1 (2007).
\bibitem{dingpla} Y. Chen, G. Rangarajan, J. Feng, and M. Ding, Phys. Lett. {\bf A 324}, 26 (2004).
\bibitem{ding-prl} M. Dhamala, G. Rangarajan, M. Ding, Phys.Rev.Lett. {\bf 100}, 18701 (2008).
\bibitem{noiprl} D. Marinazzo, M. Pellicoro, S. Stramaglia,
Phys. Rev. Lett. {\bf 100}, 144103 (2008).
\bibitem{shawe} J. Shawe-Taylor and N. Cristianini,
{\em Kernel Methods For Pattern Analysis}. (Cambridge University
Press, London, 2004).
\bibitem{nota} After a linear transformation, we may assume all the
time series to have zero mean and unit variance.
\bibitem{10fold} M. Stone, J. Roy. Stat. Soc. {\bf B 36}, 111 (1974).
\bibitem{ancona} N. Ancona and S. Stramaglia, Neural Comput. {\bf 18}, 749 (2006).
%for recent reviews see K. Hlavackova-Schindler, M. Palus, M.
%Vejmelka, J. Bhattacharya, Physics Reports {\bf 441}, 1 (2007), and
%M. Lungarella, K. Ishiguro, Y. Kuniyoshi, N. Otsu, Int. J.
%Bifurcation and Chaos {\bf 17}, 903 (2007).
\bibitem{noipla} L. Angelini, M. Pellicoro and S. Stramaglia, Phys.
Lett. {\bf A 373}, 2467 (2009).
\bibitem{smirnov}D. Smirnov and B. Bezruchko, Phys. Rev. {\bf E 79}, 046204 (2009)
\bibitem{kantz} H. Kantz, T. Schreiber, {\em Nonlinear time series analysis\/} (Cambridge University Press,
Cambridge, 1997).
\bibitem{kadanoff} L.P. Kadanoff, {\em Statistical Physics\/} (World Scientific,
Singapore, 2000).
\bibitem{papoulis}A. Papoulis, {\em Probability, Random Variables and Stochastic
Processes McGraw Hill, New York, 1965\/}.
\bibitem{boyd} S.J. Kim, K. Koh, M. Lustig, S. Boyd, D. Gorinevsky,
IEEE J. Sel. Top. in Sig. Process. {\bf 1}, 606 (2007).
\bibitem{roc} M.H. Zweig and G. Campbell, Clin. Chem. {\bf 39}, 561
(1993).
\bibitem{redundant} L. Angelini et al., Phys. Rev. {\bf E 81}, 037201
(2010).
\end{thebibliography}
\end{document}